\documentclass[12pt]{iopart}
\usepackage{wrapfig}
\usepackage[T1]{fontenc}
\usepackage{graphicx}
\usepackage{amsfonts}
\usepackage{amssymb}
\usepackage{stmaryrd}

\usepackage{cite}

\begin{document}

\title[Suppression of neutral pion production at high $p_{T}$ in heavy-ion collisions]{Suppression of neutral pion production at large transverse momentum measured with the ALICE experiment in  Pb$-$Pb collisions at $\sqrt{s_{NN}}=2.76$ TeV}

\author{G. Conesa Balbastre for the ALICE collaboration \footnote{For the full ALICE Collaboration author list and acknowledgments, see Appendix "Collaborations" of this volume.}}

\address{LPSC, Universit\'e Joseph Fourier Grenoble 1, CNRS/IN2P3, Institut Polytechnique de
Grenoble, 53 rue des Martyrs, Grenoble, France }
\ead{Gustavo.Conesa.Balbastre@cern.ch}
\begin{abstract}
The ALICE collaboration at the LHC has measured the transverse momentum spectra of neutral pions via their two
photon decay in pp and  Pb$-$Pb collisions at $\sqrt{s_{NN}}=2.76$ TeV over a broad transverse momentum range
with different subsystems: with the electromagnetic calorimeters PHOS and EMCAL and with photon conversions
in the inner material of the detectors using $e^{+}e^{-}$ pairs reconstructed with the Central Tracking System.
In this report, neutral pions production is compared between pp and  Pb$-$Pb collisions measured with conversion photons in terms of the nuclear 
modification factor, $R_{AA}$, for different centrality selections of the  Pb$-$Pb data sample. 
\end{abstract}

The study of the properties of the Quark-Gluon Plasma (QGP) is the main objective of the ultra-relativistic heavy-ion collisions 
programs at LHC, RHIC and SPS.  The QGP is a high density and hight temperature deconfined state of matter predicted by Quantum-Chromodynamics (QCD). 
Hard partons, produced in the initial stages of a heavy-ion collision, lose energy via collisional or radiative (gluon emission) mechanisms\cite{SalgWied} 
when traversing the dense and strongly interacting plasma as suggested by Bjorken~\cite{Bjorken}. One of the consequences of the parton energy loss is 
the suppression of high energy hadrons when comparing hadron production yields between heavy-ion and pp collisions. To observe this suppression and quantify 
medium effects, we calculate the $nuclear$ $modification$ $factor$ $R_{AA}$, defined as the ratio of hadron transverse momentum spectra, in this report of $\pi^{0}$,  
between  Pb$-$Pb collisions scaled by the number of binary nucleon-nucleon collisions $\langle N_{coll} \rangle$ and pp collisions, 

\begin{equation} 
R_{AA}=\frac{1}{\langle N_{coll} \rangle}\frac{(1/N^{AA}_{evt}) d^2N_{\pi^0}^{AA}/dy dp_T}{(1/N^{pp}_{evt}) d^2N_{\pi^0}^{pp}/dy dp_T},
\end{equation} 
where  the number of binary collisions $\langle N_{coll} \rangle$ is obtained from a Glauber model, as the product of the nuclear overlap function and the inelastic 
nucleon-nucleon cross section\cite{Alberica}. If no nuclear modification is present, $R_{AA}$ is unity.\\

The nuclear modification factor was measured by the ALICE \cite{ALICEppr} collaboration at LHC for inclusive charged particles produced in the transverse 
momentum range $0.1 < p_T< 100$ GeV/$c$, in  Pb$-$Pb collisions at  a center of mass energy $\sqrt{s_{NN}}=2.76$~TeV collected in 2010 \cite{ALICERaa,Jacek,Harald}.
The ALICE experiment observes a similar suppression as seen by RHIC experiments (PHENIX and STAR) \cite{RHICRaa},  but stronger ($0-5$\% centrality in 
ALICE compared to 0-10\% at RHIC, though) with a minimum $R_{AA}\sim 0.15$ ( $\sim 0.2$ at RHIC) at $p_T\simeq 6-7$ GeV/$c$.
RHIC experiments observed a different $R_{AA}$ for mesons and baryons, the measurement of different particle $R_{AA}$ is interesting at LHC to further study this different behavior.  Neutral pions are of special interest since they can be identified up to large energy, $E_{\pi^0}\sim100$ GeV in ALICE\cite{ALICEppr,EMCALppr}. 
Here we report the measurement of the neutral pions $R_{AA}$.\\

The ALICE experiment has the unique capability to measure neutral pions in the two photon decay channel, with three different subsystems, 
via the invariant mass technique: with the calorimeters  PHOS\cite{ALICEppr, PHOStdr} and EMCAL\cite{EMCALppr}, and via photon conversions
in the inner material of the detectors reconstructing $e^{+}e^{-}$ pairs  with the Central Tracking System (ITS+TPC)\cite{ConvRef}. 
 The PHOS detector is a $\mbox{PbWO}_4$ electromagnetic calorimeter that covers $\Delta \phi$ = 60$^\circ$  in azimuthal angle and $|\eta|<0.13$ 
in pseudo-rapidity, the radial distance to the  interaction point is 4.60~m. The EMCAL detector is a Pb-scintillator sampling electromagnetic calorimeter  
that covers $\Delta \phi$ = 40$^\circ$ in azimuthal angle and $|\eta| <0.7$ in 
pseudo-rapidity for the runs taken in 2010, 
and increased to  $\Delta \phi$ = 100$^\circ$ in 2011 runs, the radial distance to the interaction point is 4.28~m. 
The Inner Tracking System (ITS) and Time Projection Chamber (TPC) \cite{ALICEppr} are the main tracking devices of the ALICE experiment. 
They are located in the central barrel. The ITS consists of 6 layers equipped
with Silicon Pixel detectors detectors (SPD) positioned at radial distances of 3.9~cm and 7.6~cm, 
Silicon Drift (SDD) at 15.0~cm and 23.9~cm, and Silicon Strip detectors (SSD)
at 38.0~cm and 43.0~cm. The two innermost layers cover a pseudo-rapity
range of $|\eta|<2$ and $|\eta|<1.4$, respectively. %ITS provides precise tracking and vertex measurement.
The Time Projection Chamber (TPC) \cite{ALICEppr} is a large (85~m$^3$) 
cylindrical drift detector filled with Ne/CO$_2$/N$_2$ (85.7/9.5/4.8\%) gas
mixture. Its acceptance is $|\eta|<$0.9 over the full azimuthal angle.
The centrality of the heavy-ion collision is determined  based on the sum of the amplitudes measured by the two forward scintillator 
hodoscopes VZERO that cover  $2.8 < \eta < 5.1$ and $-3.7 < \eta < -1.7$\cite{ALICEppr} and the procedure is explained in Ref.~\cite{Alberica}. \\

\begin{figure}[htb]
\begin{center}
\includegraphics[width=7.7cm, keepaspectratio]{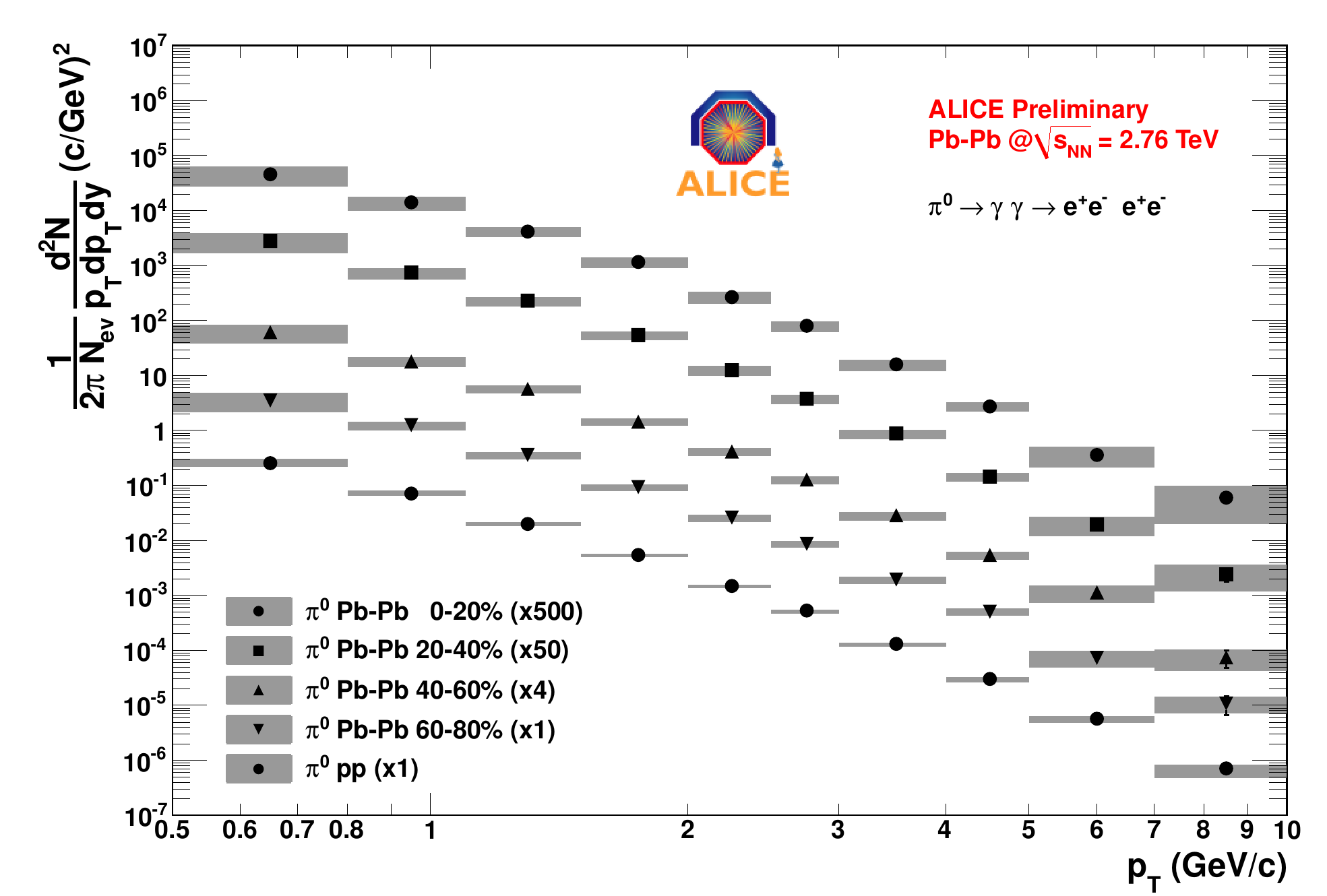}
\includegraphics[width=7.7cm, keepaspectratio]{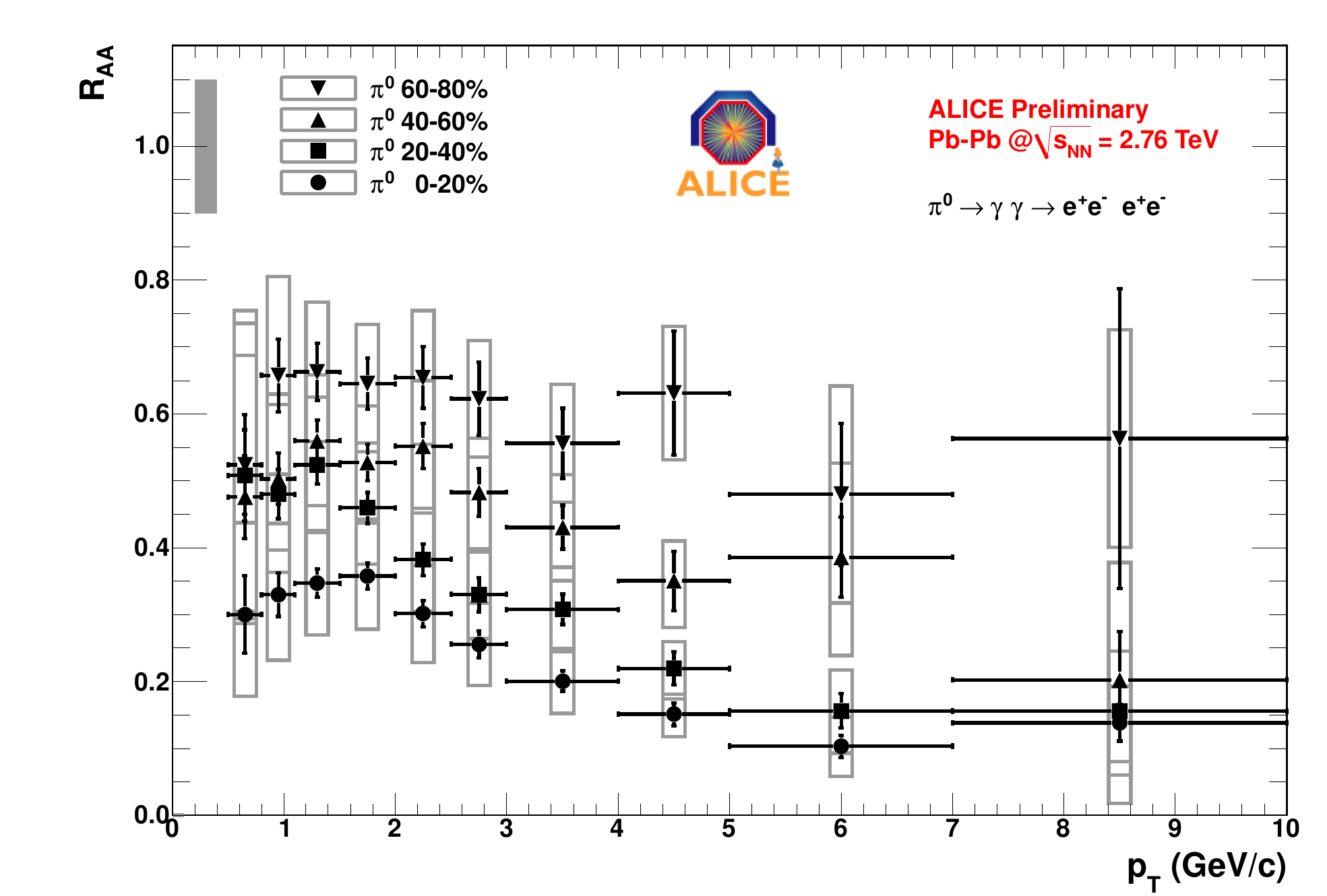}

\caption{\label{ConvSpectraRaa}{$\pi^{0}$ yield  (left frame) and nuclear modification factor (right) measured with the photon conversion technique in ALICE in  Pb$-$Pb  collisions at  $\sqrt{s_{NN}}=2.76$ TeV. Points are shown for four  Pb$-$Pb collision centralities: $0-20$\% ($\bullet$),  $20-40$\% ($\blacksquare$), $40-60$\% ($\blacktriangle$), $60-80$\% ($\blacktriangledown$). In the left plot,  Pb$-$Pb yields are scaled by different factors for visibility, pp reference points also shown in black. In both figures, boxes represent systematic errors and vertical error bars statistical errors. All points are plotted at the center of the bin.}}

\end{center}
\end{figure}

ALICE measured $\pi^0$'s  via invariant mass analysis of conversion photons in the momentum range $0.5<p_T<10$~GeV/$c$ in the  Pb$-$Pb data collected in 2010 ($\sim 1.3\times 10^7$ events with centrality between 0 and 80\% where trigger efficiency is close to 100\% and background is negligible) and also pp  data ( $\sim 5\times 10^7$ minimum bias events) collected in 2011 at $\sqrt{s_{NN}}=2.76$~TeV, as shown in Fig.~\ref{ConvSpectraRaa} (left). The conversion photons analysis allows a high resolution measurement, with a width of the invariant mass peak of about 4 MeV/$c^2$ for both pp and  Pb$-$Pb collisions at $p_T=5$~GeV/$c$ and all studied centralities.  The derived $R_{AA}$   
(Fig.~\ref{ConvSpectraRaa} (right)) shows a maximum at $p_T\sim1-2$~GeV/$c$ and decreases towards higher momenta. ALICE sees a suppression of  $\pi^0$ production at the most central events of $\sim 85\%$ for $4.5 < p_T <8.5$~GeV/$c$ and for peripheral collisions of $\sim 30-40\%$. \\

\begin{figure}[htb]
\begin{center}
\includegraphics[width=7.7cm, keepaspectratio]{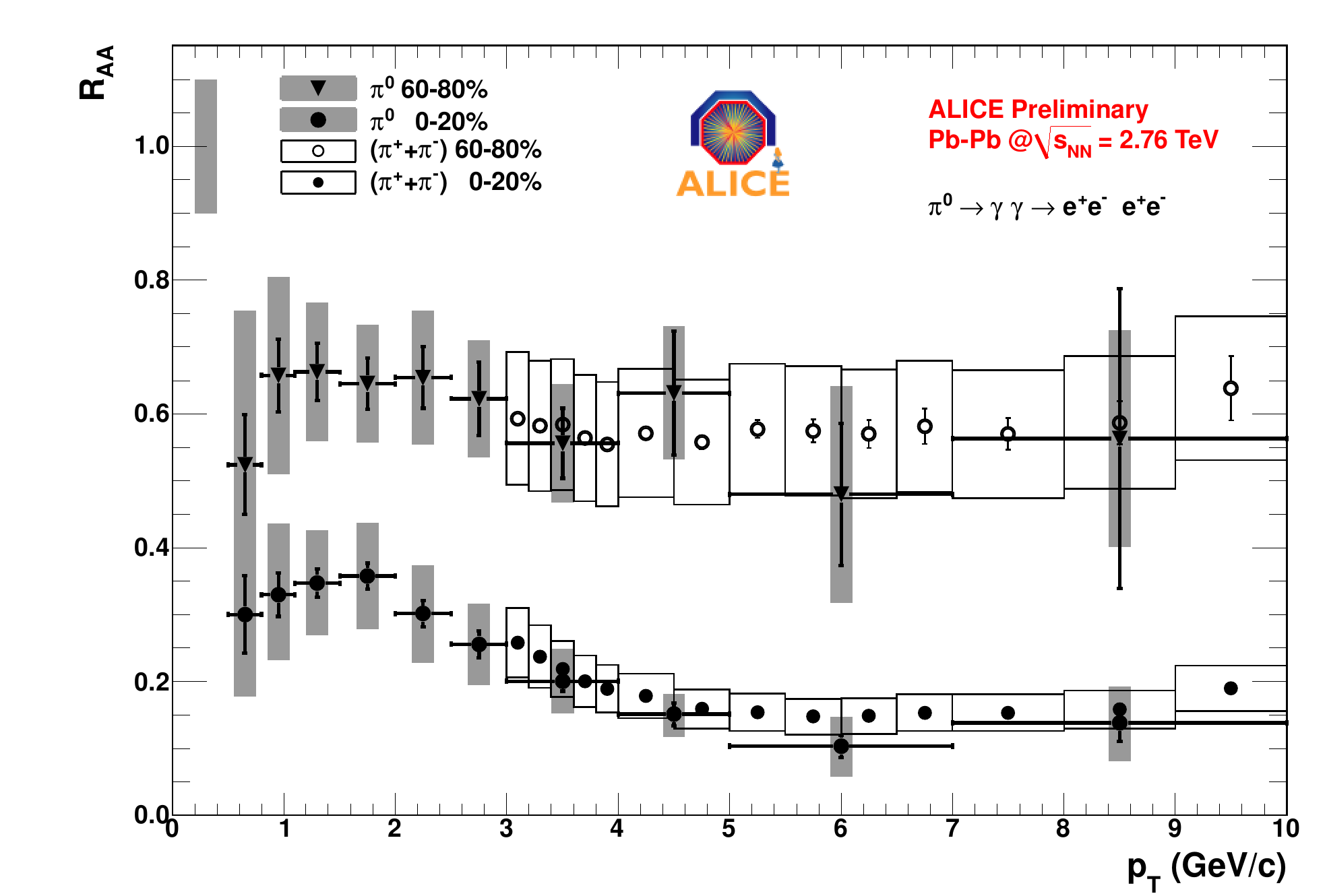}
\includegraphics[width=7.7cm, keepaspectratio]{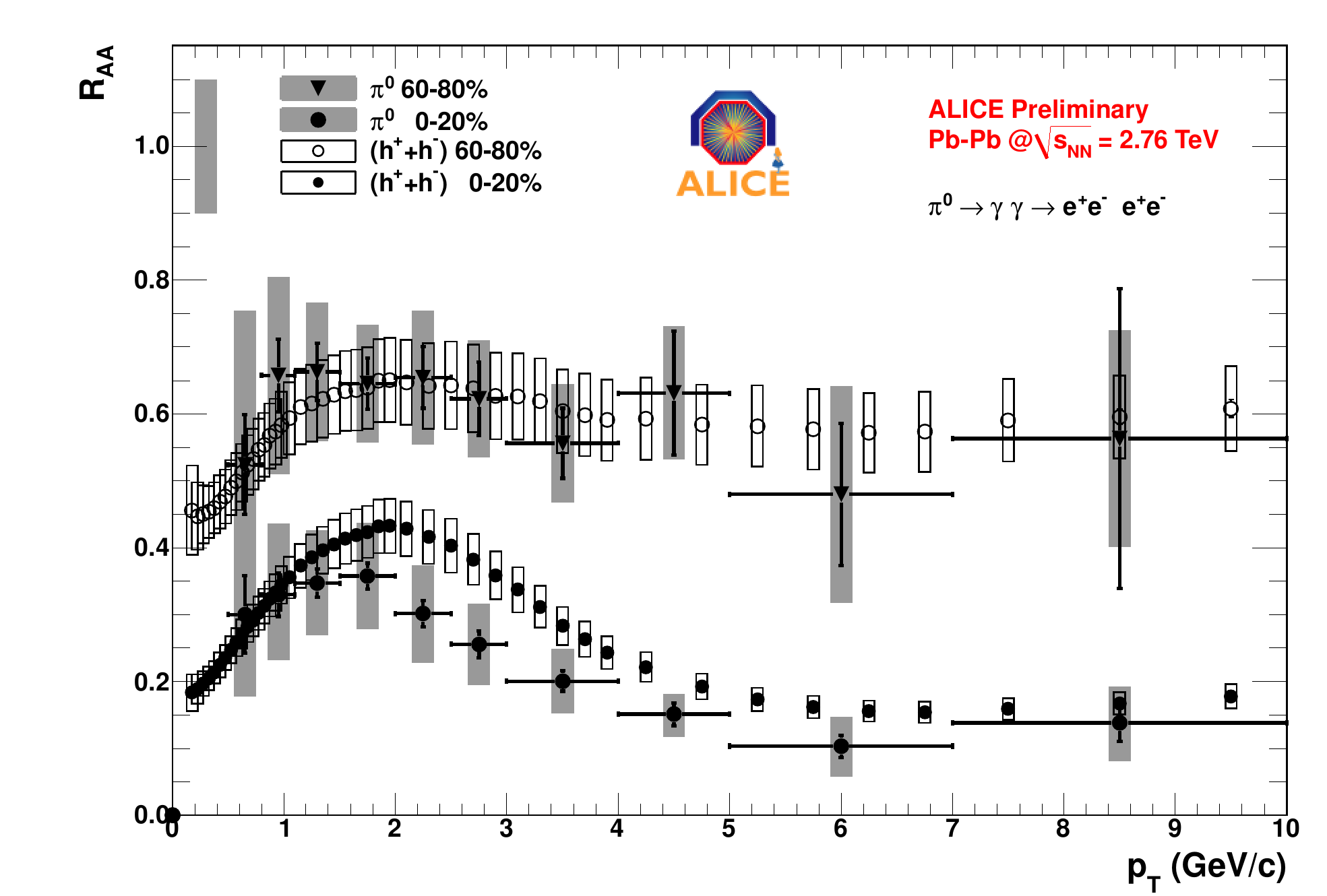}

\caption{\label{ConvRaaCompCharged}{$\pi^{0}$ nuclear modification factor measured with the conversion technique in ALICE in  Pb$-$Pb  collisions at $\sqrt{s_{NN}}=2.76$ TeV compared to that measured for charged pions (left) and inclusive charged particles (right) also in ALICE. Points are shown for two
collision centralities: central ($\bullet$) and peripheral ($\blacktriangledown$). In both figures, boxes represent systematic errors and vertical error bars statistical errors. }}

\end{center}
\end{figure}

In Figure \ref{ConvRaaCompCharged}, the $\pi^0$  $R_{AA}$ is compared to the charged pion $R_{AA}$ measured for $p_T>3$~GeV/$c$ (left frame) and charged particles 
$R_{AA}$ (right frame)\cite{Harald}. ALICE observes a good agreement within uncertainties with the charged pions measurement in all centralities. Since high momentum pions produced at mid rapidity come from hard parton fragmentation, the agreement with charged pion $R_{AA}$ was expected. This measurement extends the identified pion $R_{AA}$ measurement down to $p_T=0.5$~GeV/c. There is also agreement with charged particles $R_{AA}$ in peripheral collisions, but the shape of the $R_{AA}$ is different for more central events, which could be explained by the different particle composition and related to the observed baryon/meson anomaly at RHIC \cite{RHICRaa}.\\

 ALICE observes also $\pi^0$ transverse momentum spectra in  Pb$-$Pb collisions taken in 2010  with both the PHOS calorimeter in the range $0.6 <p_T<20$~GeV/$c$  and with the EMCAL in the range $1 <p_T<20$~GeV/$c$ at different centralities.  ALICE measured the pp reference transverse momentum spectra with PHOS and conversions technique in the short run of pp collisions at $\sqrt{s}=2.76$~TeV in minimum bias events up to about 10 GeV/$c$\cite{Klaus}. ALICE also took data in this reference run with the EMCAL trigger \cite{EMCALppr} and will allow to extend the $\pi^0$ measurement up to  20 GeV/$c$,  %(Fig. \ref{CaloRawYields}, right), 
which is crucial to have a good reference for the  $\pi^0$ $R_{AA}$ at high transverse momentum.  \\

\newpage

ALICE carried out a first measurement at the LHC of  $\pi^{0}$ yield  from central to peripheral  Pb$-$Pb collisions and pp collisions  at  $\sqrt{s}=2.76$~TeV using the photon conversion method. $\pi^{0}$ $R_{AA}$ was  measured by ALICE in the range $0.5<p_T<8.5$ GeV/$c$, showing a strong suppression of  $\pi^{0}$'s  of $85 \%$ in the range  $4.5<p_T<8.5$ GeV/$c$, a result in agreement with charged pions $R_{AA}$. The ALICE $\pi^{0}$ suppression is stronger than at RHIC where the PHENIX experiment observes a factor $4-5$ suppression for $\pi^0$'s with transverse momentum larger than 5 GeV/$c$ for a centrality of $0-10\%$ \cite{RHICRaa}.

\end{document}